\documentclass[%
 reprint,
superscriptaddress,
 amsmath,amssymb,
 aps,
 prl,
]{revtex4-2}

\usepackage{balance}
\usepackage{mathptmx}
\usepackage{graphicx} 
\usepackage{lastpage}
\usepackage{float}
\usepackage{fancyhdr}
\usepackage{fnpos}
\usepackage[english]{babel}
\usepackage{dcolumn}
\usepackage{bm}
\usepackage{xcolor}

 \usepackage{filecontents}
\usepackage{lmodern}
\usepackage{graphicx}
\usepackage{color}
\usepackage{multirow}
\usepackage{siunitx}
\usepackage{float}
\usepackage{textcomp,gensymb}
\usepackage{mathrsfs}
\usepackage{bm}
\usepackage{relsize} 
\usepackage[hidelinks]{hyperref}
\usepackage{xr}
\usepackage{booktabs}

\graphicspath{{figures/}}

\definecolor{myblue}{rgb}{0.0, 0.0, 0.45}
\newcommand{\sIsIII}[1]{\textcolor{myblue}{#1}}

\definecolor{myred}{rgb}{0.5, 0.0, 0.0}
\newcommand{\sIsII}[1]{\textcolor{myred}{#1}}

\definecolor{mygreen}{rgb}{0.0, 0.5, 0.0}
\newcommand{\sIIsIII}[1]{\textcolor{mygreen}{#1}}

\newcommand{\sref}[1]{S\ref{#1}}

\makeatletter
\newcommand*{\addFileDependency}[1]{
  \typeout{(#1)}
  \@addtofilelist{#1}
  \IfFileExists{#1}{}{\typeout{No file #1.}}
}
\makeatother

\newcommand*{\myexternaldocument}[1]{%
    \externaldocument[SM-]{#1}%
    \addFileDependency{#1.tex}%
    \addFileDependency{#1.aux}%
}
	
\myexternaldocument{arxivcghydrogelSI}

\begin{document}

\title{Nanostars planarity modulates the elasticity of DNA hydrogels} 

\author{Yair Augusto Gutierrez Fosado}
\affiliation{School of Physics and Astronomy, University of Edinburgh, Peter Guthrie Tait Road, Edinburgh, EH9 3FD, UK\\ For correspondence: yair.fosado@ed.ac.uk}

\begin{abstract}
In analogy with classic rigidity problems of networks and frames, the elastic properties of hydrogels made of DNA nanostars (DNAns) are expected to strongly depend on the precise geometry of their building blocks. However, it is currently not possible to determine DNAns shape experimentally. Computational coarse-grained models that can retain the correct geometry of DNA nanostars and account for the bulk properties observed in recent experiments could provide missing insights. In this study, we perform metadynamics simulations to obtain the preferred configuration of three-armed DNA nanostars simulated with the oxDNA model. Based on these results we introduce a coarse-grained computational model of nanostars that can self assemble into complex three dimensional percolating networks. We compare two systems with different designs, in which either planar or non-planar nanostars are used. Structural and network analysis reveal completely different features for the two cases, leading to two contrasting elastic properties. The mobility of molecules is larger in the non-planar case, which is consistent with a lower viscosity measured from Green-Kubo simulations in equilibrium. To the best of our knowledge, this is the first work connecting the geometry of DNAns with the bulk rheological properties of DNA hydrogels and may inform the design of future DNA based materials.
\end{abstract}

\maketitle

\section{Introduction}
The specific binding rules through which DNA nucleotides form pairs (Adenine-Thymine and Guanine-Cytosine) has been long known to be key for the storage and replication of the genetic information. This same mechanism is employed in DNA nanotechnology to form DNA motifs, i.e., artificial structures with high programmability. Here we study DNA nanostars (ns), motifs made by several double-stranded (ds) arms connected into a single structure~\cite{Hydrogelassemble2006,C6QM00176A}. The number of arms per DNAns defines its valence ($f$). Each of these arms is provided with a sticky end which, under appropriate conditions, allows nanostars to hybridize into complex three dimensional percolating networks to form a DNA hydrogel~\cite{Sciortino2013}. It is due to their inherent biocompatibility, besides the possibility of functionalization~\cite{Morya2020} and systematic control over their mechanical properties~\cite{Luo2012,Bush2021,LDiM019}, that hydrogels have emerged as promising materials in the development of diverse applications such as biosensing~\cite{LIU20181}, drug delivery~\cite{NISHIKAWA2011488} and tissue engineering~\cite{tissue}, among others.

In recent years, there have been very successful studies to characterize the properties of DNA hydrogels. In reference~\cite{Smallenburg2013}, for example, it was suggested that key components for the formation of these materials were the limited valence~\cite{Patchycolloidslimitval,Biffieqgelslowv} and the internal flexibility of DNAns~\cite{OmarA2017}. The phase diagram of gels made of DNAns with $f=3$ is reported in ~\cite{Sciortino2013}. In reference~\cite{Eiser2018}, microrheology experiments of threevalent DNAns were performed at very large concentrations (20 $mg/mL$), at which the system exhibits a phase transition from a fluid of disconnected DNAns, at high temperatures, to a fully bonded state with maximum network stiffness at low temperature. It was also proved that a flexible section in the vicinity of the sticky ends (as conferred by unpaired bases in the DNAns design) produces gels with lower bulk elasticity. The role of valence was tested in \cite{Conrad2019} by oscillatory bulk rheology of gels made from DNAns designs with different number of dsDNA arms (ranging from three to six). It was found that at the same concentration of DNAns, the higher the nanostar valence the stiffer the network formed. Finally, viscoelastic properties of tetravalent DNAns at different salt concentration were investigated in reference~\cite{Jeon2018}. 

The previous results expose an important aspect of DNA hydrogels: that beyond base-pairing thermodynamics, the precise topology of DNAns is a key component that determines the elastic properties of networks~\cite{Zhou19119primaryloops,Gu2018networktopology,Jeremiah2019}. Understanding the relationship between the geometry of nanostars and the macroscopic mechanical properties of DNA hydrogels is essential and a subject of ongoing study. Computer simulations can shed light into this aspect. Indeed, simulations with the oxDNA~\cite{Ouldridge:2011} model of a system of tetravalent DNAns have been performed in the past~\cite{Rovigatti2014,RovigattiGelsNever,rovigatti2017mixture,Eiser2020FJ}. These simulations yield to a phase diagram in fair agreement with experimental results~\cite{Sciortino2013} and provided the first strong evidence that gels of DNAns should not crystallize. However, the level of coarse-graining and specificity in the oxDNA model, would make computationally unfeasible to simulate a system comprising more than a few hundreds nanostars, not to mention, exploring the role of different geometrical designs. Models adopting a lower resolution are therefore needed, yet they need to be inferred from higher resolution ones.

Recently, a bead-spring coarse-grained model of trivalent DNAns was introduced in reference~\cite{Xing2019}. In this model, each nano star is represented by ten beads arranged into a Y-shaped planar structure. Adjacent nanostars can hybridized through specific binding sites, capturing in this way the overall network formation. The model represents an important step in the study of thermodynamic, structural and rheological features of percolating networks. However, since experiments cannot resolve the detailed geometry of nanostars, their shape is usually assumed to be planar in this type of simulations.

Here, we propose an innovative way of inferring the geometry of a single DNAns from metadynamics simulations with higher resolution models. This biasing technique allows the quick inspection of DNAns conformations by flattening the free energy landscape of the system. From this computation we characterize and tune how the planarity of DNAns changes in the equilibrium DNAns conformation by introducing slight modifications in their design. We then build a coarse-grain model that accounts for the correct geometry of DNAns and we compare the structural features, melting response and linear elasticity of two networks made of either, planar or non-planar molecules.

\begin{figure}[t]
\centering
\includegraphics[width=0.49\textwidth]{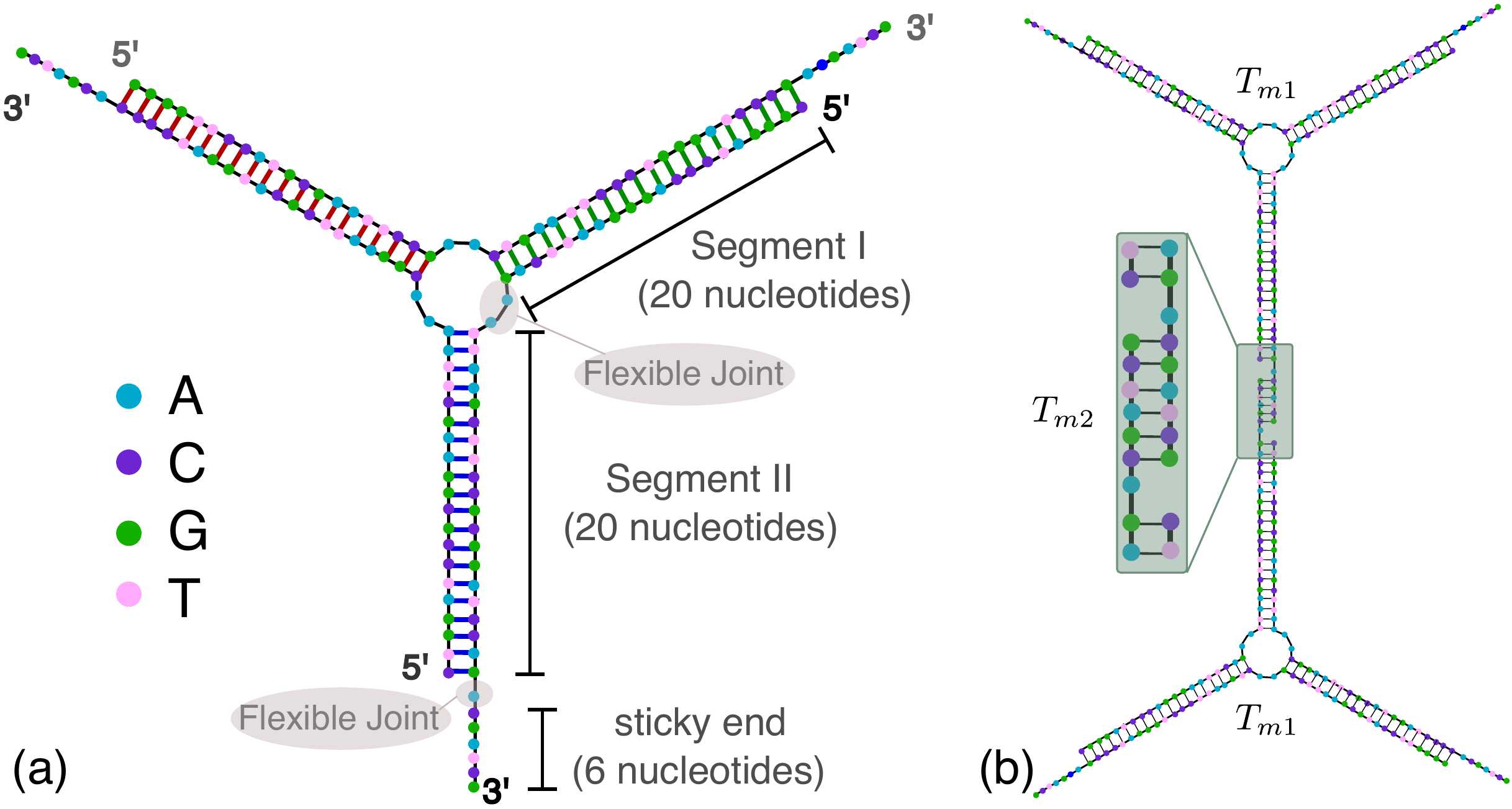}
\caption{DNA nanostars design. (a) schematic representation of the ssDNA oligonucleotides assemble into a DNAns. Different nucleotides are identified through a colour code: Adenine (A) cyan, Cytosine (C) purple, Guanine (G) green and Thymine (T) in pink. The directionality of the chains, from $5'$ to $3'$, is indicated in the first and last nucleotides of each ssDNA. Labels of the five functional parts described in Table\ref{table:seq} (two flexible joints, the sticky end and the two segments) are shown for one of the strands. Backbones are depicted in black and hydrogen bonds are coloured according to Table\ref{table:seq}. (b) Cartoon of the hybridization between two DNAns connected through the sticky ends. Sequences used here ensure that the melting temperature of the network ($T_{m2}$) is lower than the melting of individual stars ($T_{m1}$).}
\label{fig:design}
\end{figure}

\section{\label{sec.DNAdesign} Nanostars design}
The oxDNA is a well established single nucleotide resolution coarse-grained model, that is based on force fields tuned to account for several geometrical and thermodynamic features of single and double stranded DNA. Here we use its most recent implementation~\cite{oxDNALAMMPS} into the LAMMPS~\cite{LAMMPS} engine, in order to simulate DNA nanostars made of three single-stranded (ss) oligonucleotides. Sequences are reported in table \ref{table:seq} and are similar to the ones used in reference~\cite{Conrad2019}. Each ssDNA is 49 bases long and consist of five regions. The segments I and II (20 nucleotides long each) are designed to form the three dsDNA arms. In between the two segments there are two A-nucleotides acting as a spacer and forming the flexible joint at the nanostar core, (FJC). The sticky end is formed by 6 bases and has the same sequence for the three oligonucleotides, allowing in this way the non-specific hybridization of two nanostars: any of the three arms of one ns can hybridize with any (but only one) of the arms of another ns. Finally, there is a second flexible joint (FJ), formed by an A-nucleotide in between segment II and the sticky end. Figures~\ref{fig:design}(a)-(b) show schematic representations of the DNAns assemble and the binding of two nanostars, respectively. Beyond structural features, the overall sequences employed provide certain stability. The melting temperature ($T_{m1}$) of individual stars is larger than the melting temperature ($T_{m2}$) of the hybridization between stars. This ensures that there is a range of temperatures (below $T_{m1}$) in which the core structure of nanostars does not suffer major changes and the assemble/break-up of the network can be proved. This is the regime considered in the present work.
 
\begin{table}[h]
\centering
{\Huge
\resizebox{\columnwidth}{!}
{
\begin{tabular}{*5c}
\toprule
Segment I & FJC & Segment II & FJ & Sticky end    \\
\midrule
 \begin{tabular}{@{\hspace{0cm}}c@{\hspace{0cm}}}
 $5^{\prime}-$  \sIsIII{CTGGATCCGCGGAAGCTTAA}\\
 $5^{\prime}-$   \sIsII{GGGGATCCATGCGAATTCCG}\\
 $5^{\prime}-$ \sIIsIII{CGGGATCCCAGGGAATTCAG}
 \end{tabular} 
 & \begin{tabular}{@{\hspace{0cm}}c@{\hspace{0cm}}}
 AA \\
 AA \\
 AA
 \end{tabular}
 & \begin{tabular}{@{\hspace{0cm}}c@{\hspace{0cm}}}
   \sIsII{CGGAATTCGCATGGATCCCC} \\
 \sIIsIII{CTGAATTCCCTGGGATCCCG} \\
  \sIsIII{TTAAGCTTCCGCGGATCCAG}
 \end{tabular}
 & \begin{tabular}{@{\hspace{0cm}}c@{\hspace{0cm}}}
 A \\
 A \\
 A
 \end{tabular}
 & \begin{tabular}{@{\hspace{0cm}}c@{\hspace{0cm}}}
 CGATCG $-3^{\prime}$ \\ 
 CGATCG $-3^{\prime}$ \\   
 CGATCG $-3^{\prime}$ 
 \end{tabular}\\
\bottomrule
\end{tabular}
}
}
\caption{Strand sequence used in the nano star design with valence $f=3$ and $n=2$ unpaired nucleotides at FJC. Each row represents a different ssDNA oligonucleotide. Segments with the same colour have complementary sequences to form the double stranded arms as shown in Fig.~\ref{fig:design}.}
\label{table:seq}
\end{table}

\begin{figure*}[ht]
\centering
\includegraphics[width=1.0\textwidth]{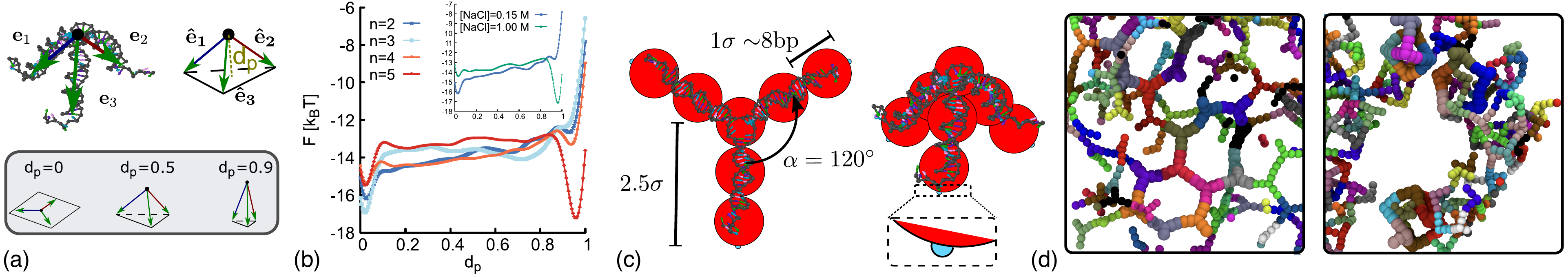}
\caption{Coarse-grain modelling of DNA nanostars. (\textbf{a}) Schematic representation of $d_{p}$. Top panel shows the vectors $\mathbf{e}_{1}, \mathbf{e}_{2}, \mathbf{e}_{3}$ pointing from the core of the molecule (black dot) to the last base-pair in each arm. The variable $d_{p}$ represents the distance from the core of the molecule to the plane touching the end of the normalized vectors $\mathbf{\hat{e}}_{1}, \mathbf{\hat{e}}_{2}, \mathbf{\hat{e}}_{3}$ (see ESI for details). Bottom panel shows sketches of three DNAns with different degree of planarity, from let to right the value of $d_{p}$ is 0, 0.6 and 0.9. (\textbf{b}) Free Energy Landscape as a function of $d_{p}$ for DNAns simulated with the oxDNA model at [NaCl]=0.15 M. Different colours represent results for different number of unpaired nucleotides at the FJC. Inset shows comparison for $n=2$ at two salt concentrations. (\textbf{c}) Coarse-grain geometries of planar (left, with $d_{p}=0$) and non-planar (right, with $d_{p}=0.6$) molecules. Red beads represent the dsDNA sections of the molecules. Small patches at the end of each arm are displayed in cyan. (\textbf{d}) Snapshots from simulations of networks formed when using planar (left) and non-planar (right) DNAns.}
\label{fig:fel}
\end{figure*}

\section{\label{sec.meta}Planarity of DNAns from Metadynamics simulations} 
Molecular dynamics (MD) simulations are often used to assist the design of DNA nanostructures. However, in scenarios where the free energy landscape (FEL) of the system is complex, with several local minima separated by large energetic barriers, it becomes difficult to ensure that in the course of the simulation the phase-space to move from one minimum to the next one has been completely explored. Metadynamics~\cite{2002Laio,2008Laio} is a computational method that provides a framework to determine free energies and accelerate rare events, allowing the system to escape from local minima in the FEL. In essence, in the metadynamics simulations we need to find a set of collective variables (CVs), $\zeta_{i}(\mathbf{r})$, that gives relevant information about the state of the system and that it only depends on the position, $\mathbf{r}$, of particles. Then, the system is biased to explore different regions on the phase-space by adding a history-dependent Gaussian potential $U_{G}(\zeta_{i}(\mathbf{r}),t)$. The basic assumption of metadynamics is that after a sufficiently long time, $U_{G}$ provides an estimate of the free energy landscape $F(\zeta_{i})$. A more detailed explanation of the method is provided in the ESI.

The method described above is used here to obtain the FEL as function of $d_{p}$, a collective variable related to the degree of planarity of a DNAns. This variable is defined as the magnitude of the vector pointing from the core of the molecule to the plane touching the tip of the three unitary vectors representing the direction of the three dsDNA arms (see Fig.\ref{fig:fel}(a) and ESI). The lower the value of $d_p$ the more planar the molecule. Figure~\ref{fig:fel}(b) shows the FEL of DNAns designs with a varying number ($n$) of unpaired nucleotides at the core of the molecule, when the salt concentration of the system is [NaCl]=0.15 M. For $n=2$ there is a global minimum located at $d_{p}\sim 0$ (planar), with an energetic barrier to overcome before exploring other regions. This indicates that the planar configuration of the nanostars would be the most favourable. As $n$ increases, the energy barrier becomes lower and more local minima are developed around $d_{p}=0$ and 0.7. In these cases it is expected that the planarity of the molecule fluctuates strongly over time. For $n=5$ the free energy exhibits two clear minima, with the global one favouring the non-planar configuration of the nanostar. These results can be explained by considering that for small values of $n$, each dsDNA arm is constrained by the presence of the other two and non-planar configurations will create stress, particularly at the core of the molecule. As $n$ increases, the dsDNA arms become more disconnected and more configurations are accessible.

In the inset of Fig.~\ref{fig:fel}(b) we compare the FEL of DNAns with $n=2$ at two different salt concentrations. The position of the global minimum changes from $d_{p}\sim 0$ (planar) to $d_{p}\sim 1$ (non-planar) as [NaCl] changes from 0.15 M to 1 M. This behaviour occurs because at high salt concentrations the electrostatic repulsion within the DNAns is screened. Therefore, the effective diameter of the dsDNA arms decreases~\cite{diametersalt}, facilitating the non-planar configuration of the DNAns.

\begin{figure*}[t]
\centering
\includegraphics[width=0.96\textwidth]{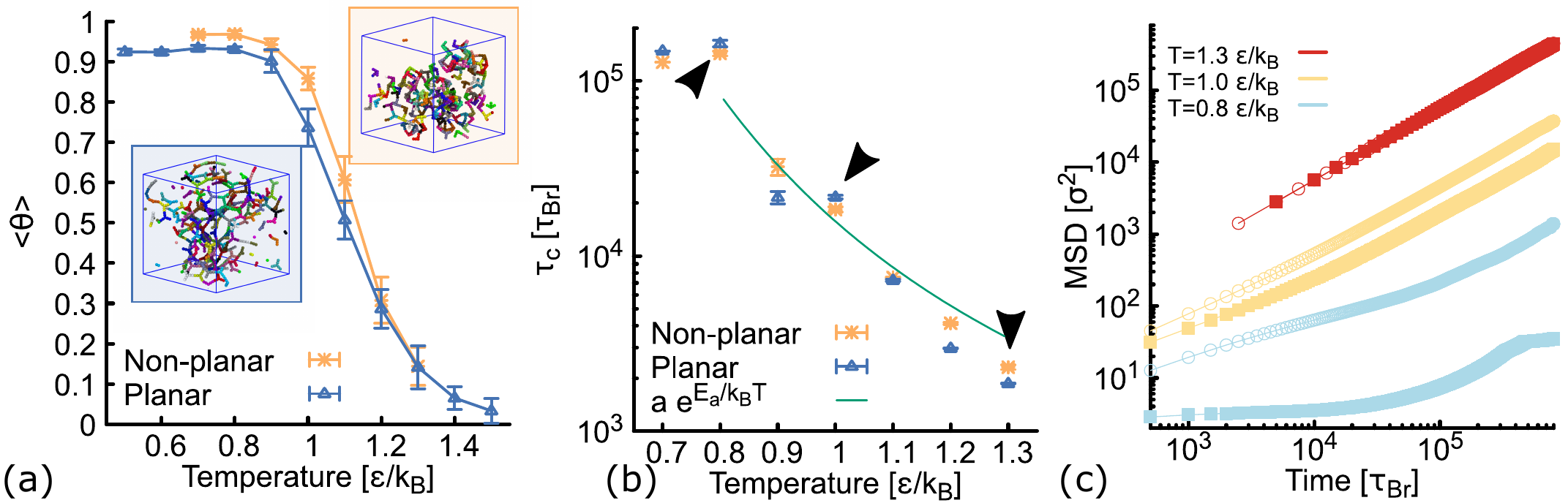}
\caption{Results from simulations at different temperatures of systems comprising either planar (blue) or non-planar (orange) molecules. (\textbf{a}) Melting curves obtained via MD simulations from the averaged number of connected patches at long times, (see Fig.~\sref{SM-fig:changeT} and ESI). Note that the melting temperature ($T\sim1.15 \epsilon/k_{B}$) of non-planar molecules is larger than the one ($T=1.1\epsilon/k_{B}$) for planar molecules. Insets show snapshots of typical configurations obtained in simulations. (\textbf{b}) Log-linear plot of the relaxation time of the networks, obtained from the autocorrelation function of $N_{c}(t)$ (see Fig.~\sref{SM-fig:changeT}). The green line represents the fit to the data using an exponential function with the general form $ae^{E_{a}/k_{B}T}$. From the fit we obtain $a=0.8\tau_{Br}$ and $E_{a}=9.7$ $k_{B}T$, which is compatible with the depth of the Morse attractive potential set in simulations (see ESI). Black arrows indicate the temperatures related to the MSD shown in the next panel. (\textbf{c}) Log-Log plot of the MSD of DNAns for planar (filled squares) and non planar (open circles).  Colours represent results at different temperatures: below ($T=0.8~\epsilon/k_{B}$ in blue), close ($T=1.0~\epsilon/k_{B}$ in yellow) and above ($T=1.3~\epsilon/k_{B}$ in red) the melting temperature.}
\label{fig:meltandtc}
\end{figure*}

The previous results suggest that minor changes in the design of the DNAns have major implications on their shape. In particular, unpaired bases at the core and changes in salt concentration affect their planarity. To investigate how this change impact bulk gel properties, here we introduce a corse-grained model of trivalent nanostars with different geometry (Fig.\ref{fig:fel}(c)). For simplicity, each nanostar is modelled as a rigid body made of seven beads (depicted in red) that represent the core of the molecule and the three dsDNA arms. Attractive patches (depicted in cyan) are placed at the edge of the last bead in each arm, mimicking in this way the sticky ends interactions. Beads have an excluded volume of $\sigma=2.5$ nm $\sim 8$ bp, so nanostars cannot overlap. Patches interaction is set via a Morse potential with energy $\epsilon_{m}$ that ensures the attraction of patches in a radius of 0.2 $\sigma$ (see ESI for details). In the planar case ($d_{p}=0$), the angle between consecutive arms is $\alpha=120\degree$, the average value found in simulations. In the non-planar case we use and intermediate value of $d_{p}=0.6$, the geometry in the coarse-grain model is the one of a tetrahedron with an equilateral triangle base and three equal isosceles triangle sides. In the ESI we show that results are consistent across different values of $d_{p}$. Figure~\ref{fig:fel}(d) shows snapshots from simulations of networks formed when using planar and non-planar molecules. Clear differences in the shape and connectivity of the networks can be seen. In the following sections we investigate more in detail these differences and how they affect the elastic properties of the networks.

\section{Melting curves and relaxation time}
We first study the formation of the network via molecular dynamic simulations employing the model previously described (see ESI for more details on the model and the MD). In the simulations reported here, we start from an equilibrated configuration of $N=175$ unconnected nanostars (with $\epsilon_{m}=0$) at temperature $T$. The system is in a cubic box of size  $L=40~\sigma$ such as the volume fraction is $\rho=0.01$. Then, we turn on the morse attraction between patches, and record the time evolution of the system until a steady state is reached.

An observable that can be directly compared with experiments is the fraction of connected DNAns, $\vartheta=2N_{c}(t)/Nf$, with $N_{c}(t)$ the total number of contacts between patches at time $t$. The plot of the equilibrium value $< \vartheta >$ as a function of temperature can be identified with the melting curve of the system. This is reported in Fig.~\ref{fig:meltandtc}(a) for networks formed with planar and non-planar molecules. At high temperatures $< \vartheta > \to 0$, indicating that patches are dissociated and resembling a gaseous state for the two systems. As temperature decreases, DNAns bonds start to form, but the melting temperature (at which $<\vartheta>=0.5$) is larger for the network made of non-planar molecules. At low temperatures $<\vartheta>$ plateaus at a value close to 1 for both systems. The networks have formed all the possible bonds, but the non-planar molecules present a consistent higher fraction of connection. As we will show in the next section, this effect is related to the geometry of DNAns (see Fig.~\ref{fig:structure}(a)).

The characteristic time ($\tau_{c}$) for network reconfiguration~\cite{Sciortino2017PRL}, i.e., the time that it takes for one of the DNAns to unbind and bind somewhere else, can be measured from the autocorrelation function of $N_{c}(t)$ (see ESI). Values obtained at different temperatures are reported in Fig.~\ref{fig:meltandtc}(b). As expected, at high temperatures thermal fluctuations facilitate the unbinding of patches and in consequence $\tau_{c}$ is small. At low temperatures instead, thermal fluctuations are weaker and therefore, the relaxation time of the network increases. We note that $\tau_{c}$ plateaus at $T\leq0.8$ $\epsilon/k_{B}$, indicating that fluctuations of $N_{c}(t)$ in this range of temperatures are very similar.

It has been shown~\cite{Conrad2019} that $\tau_{c}$ exhibits an Arrhenius dependence: $\tau_{c}\propto e^{E_{a}/k_{B}T}$ where $k_{B}$ is the Boltzmann constant and $E_{a}$ is associated to the binding energy of the sticky end. A fit to our data using this equation is depicted by the green line in Fig.~\ref{fig:meltandtc}(b). The agreement is reasonable, considering the simplicity of our model. Remarkably, at low temperature $\tau_{c}$ is larger for the planar case and, as the temperature increases, this difference becomes smaller. The mean squared displacement (MSD) of nanostars (shown in Fig.~\ref{fig:meltandtc}(c) for three different temperatures) is in agreement with the previous result. At $T=0.8~\epsilon/k_{B}$, the network made of non-planar molecules shows a larger mobility. The difference in mobility decreases at $T=1.0~\epsilon/k_{B}$ and it becomes negligible at $T=1.3~\epsilon/k_{B}$, when both systems are fully disconnected.

\section{Structural analysis}
\begin{figure*}[t]
\centering
\includegraphics[width=1.0\textwidth]{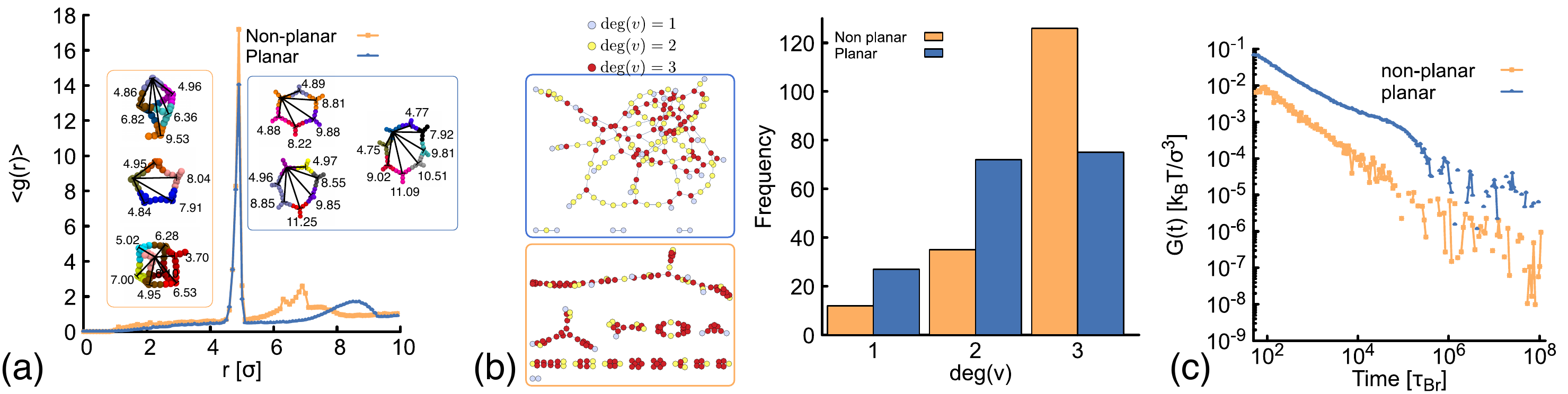}
\caption{Network structure of planar (blue) and non-planar (orange) molecules. All the data shown here correspond to simulations performed at $\rho=0.01$ (\textbf{a}) Radial distribution function computed from simulations at $T=1.0 \epsilon/k_{B}$ (details and results at different temperatures can be found in the ESI and Fig.~\sref{SM-fig:changeT}). In the insets, typical geometries found in simulations of the two networks are shown. In each "ring" structure, black lines connect the core of two DNAns and the labels show the distance (in simulation units) between them. The top-most ring shown inside the orange and blue rectangles, correspond to rings made of six DNAns in the non-planar and planar cases. While in the former there are eight contacts between neighbouring DNAns, in the latter there are only six contacts. (\textbf{b}) Network diagrams (left) and histograms (right) showing the connectivity between nanostars. (\textbf{c}) Autocorrelation ($G(t)$) of the stress-tensor.}
\label{fig:structure}
\end{figure*}

In order to understand the structure of the networks formed in our simulations, we first compute the radial distribution function (RDF), $g(r)$, using the position of the beads at the core of the molecules and averaging over configurations in the steady state (see ESI for details). Results are depicted in Fig.~\ref{fig:structure}(a). Both networks, made of planar and non-planar molecules show a global maximum of $g(r)$ located close to $r=5$ $\sigma$. This corresponds to the average distance between the cores of two bound nanostars. A bump is located in between $7.5< r < 9$ $\sigma$ in the planar case, with local maximum at $r=8.5$ $\sigma$ (distance between second nearest neighbours). By contrast, in the non-planar case there are two distinct local maxima located at $r=6.4$ and 6.9 $\sigma$, corresponding to the distance between second and third nearest neighbours, respectively. These results are consistent with the ring structures observed in simulations and depicted in the insets of  Fig.~\ref{fig:structure}(a). Remarkably, simulations displayed a rich variety of unanticipated structures. For example, while rings made of only six DNAns are expected in the planar case (because $\alpha_{0}=120\degree$), we found some rings made of seven or even eight DNAns. In the case of non-planar molecules, not only rings are formed but also box-like structures. It is worth noting here that if we compare, for example, rings made of six DNAns in the insets of Fig.~\ref{fig:structure}(a), eight contacts are made between neighbouring DNAns in the non-planar case and only six contacts in the planar case. This result explains why $<\vartheta>$ is consistently larger in Fig.\ref{fig:meltandtc}(a) and suggests that the non-planarity of the molecules would affect the degree of connectivity of DNAns in the network as we will see below.

In Fig.~\ref{fig:structure}(b) we show diagrams of the connection between DNAns in the network. In this network diagrams, each DNA nanostar is represented by a circle (also called vertex). A line (also called edge) is drawn between any two connected nanostars and colours are used to represent the degree of a vertex (deg$(v)$), i.e., the number of DNA nanostars connected to that vertex. Because the DNAns valence is $f=3$, the value of deg$(v)$ can be either: 0 (for isolated stars, not shown in the plot), 1 (light-blue circles), 2 (yellow circles) or 3 (red circles). The histogram showing the frequency of the nanostars with certain degree of connection is also shown. As it can be seen, the number of DNAns fully connected (deg$(v)=3$) is larger for the non-planar molecules. This is also reflected in the higher density of red circles in the network diagram at the bottom, which would explain why $<\vartheta>$ in Fig.~\ref{fig:meltandtc}(a) is smaller for the planar network.

In the network diagrams, a connected component is a set of vertices with edges spanning paths to connect any two of them. The larger the set of vertices in a component, the higher the degree of connectivity in the system. By inspecting the network diagrams in Fig.~\ref{fig:structure}(b), it is evident that in the planar system most of the DNAns participate in the network and a few of them form small clusters. On the other hand, the non-planar system shows several clusters. Therefore, the degree of connectivity is larger for the planar case. One way to show this, is by computing the number ($c_{s}$) of DNAns that are part of the largest component of the network (normalized by the total number of DNAns, $N$). This is shown in the ESI Fig.~\sref{SM-fig:cluster} for different temperatures. Results are consistent across the range of temperatures explored here. The implications of these observation on the elastic behaviour of the networks are explored in the following section.

\section{Viscosity}
Here we compute the zero shear viscosity ($\eta$) of the network using the Green-Kubo~\cite{greenkubo} relations. In Fig.\ref{fig:structure}(c) the autocorrelation ($G(t)$) of the off-diagonal components of the stress-tensor is shown. This was computed using the multiple-tau correlator method~\cite{taucorr} from long equilibrium MD simulations (see ESI). The viscosity of the system is then obtained as the integral of $G(t)$. The network formed by planar molecules has a viscosity ($\eta_{p}=573$ $ k_{B}T\tau_{Br}/\sigma^{3} = 5.5 Pas$). As comparison, the viscosity of gels at room temperature and low salt concentration, made of tetravalent DNAns and at a larger concentration of [DNAns]=220 $\mu$M, is $\eta \sim 100 Pas$~\cite{viscositytetravalent}. Considering that the latter two conditions increase the viscosity of the system, the simplicity of our model and the smaller volume fraction used here, our estimate of the viscosity is reasonably. Remarkably, the viscosity when the network is formed by non-planar molecules ($\eta_{np}=29$ $k_{B}T\tau_{Br}/\sigma^{3}=0.28 Pas$) is twenty times lower than the one made by planar nanostars. In the ESI we show that the net decrease in viscosity observed in simulations would depend on the degree of non-planarity of the molecule. The larger the value of $d_{p}$, the smaller the viscosity of the network.

\section{Conclusions}
In summary, we have introduced a method to infer the geometry of DNA nanostars from metadynamics simulations. We found a way to regulate the planarity of DNAns by varying the number of unpaired nucleotides at the core. Our simulations provide a rich physical insight on how the geometry of DNAns has a major impact on the connectivity of the network and ultimately on the viscosity of the DNA hydrogels. We also showed that a different mechanism to control the planarity of DNAns is by increasing the salt concentration of the system. However, we anticipate that in these conditions more variables should be considered. The salt concentration would not only modulate the shape of nanostars, but would also make sticky-ends hybridization more stable, increasing the time for network reconfiguration $\tau_{c}$. 

While our coarse-grained model is currently less sophisticated than other mesoscopic models, such as oxDNA, it is also robust enough to capture the overall formation of the network and computationally efficient to probe properties at large volume fractions if desired. Importantly, this model can also be extended to treat DNAns as bead-spring polymers (not longer rigid bodies) and to include fluctuations in the geometry of DNAns that are expected for some DNAns designs, according to the FEL obtained here.

\section*{Acknowledgements}
This project has received funding from the European Research Council (ERC) under the European Union’s Horizon 2020 research and innovation programme (grant agreement No 947918, TAP). I would like to thank D. Michieletto and G. Palombo for the fruitful discussions.


\bibliographystyle{apsrev4-1}
\bibliography{cghydrogel}

\makeatletter\@input{intermediate1.tex}\makeatother
\makeatletter\@input{intermediate2.tex}\makeatother

\end{document}


\title{Supplementary Information: Planarity modulates the elasticity of DNA hydrogels} 

\author{Yair Augusto Gutierrez Fosado}
\affiliation{School of Physics and Astronomy, University of Edinburgh, Peter Guthrie Tait Road, Edinburgh, EH9 3FD, UK\\ For correspondence: yair.fosado@ed.ac.uk}

\maketitle

\tableofcontents
\setcounter{secnumdepth}{1}
\clearpage

\section{\label{App.detailsmeta}Details of Metadynamics simulations}
\subsection{Reminder of the theory}
As described in the main text, Metadynamics (MetaD) is a computer simulation method that can be used to estimate the free energy landscape (FEL) of a system. Importantly, it is assumed that this FEL can be written in terms of few Collective variables (CVs), $\zeta_{i}(\mathbf{r})$, that depend only on the positions, ($\mathbf{r}$), of particles and that can distinguish between any two (or more) states of the system. Different states are represented by minima in the FEL. 

In MetaD, the sampling of the phase-space of a system is facilitated by the introduction of a history-dependent bias potential that forces the system to migrate from one minimum of the FEL to the next one. When just one CV is considered the potential, constructed as a sum of Gaussians at time $t$, has the form:

\begin{equation}
U_{G}(\zeta(\mathbf{r}),t) = \displaystyle \sum_{t^{\prime}=\tau_{G}, 2\tau_{G},...} B \mathrm{exp} \left(- \frac{[\zeta(\mathbf{r}) - \nu(t^{\prime})]^{2}}{2\Delta^{2}} \right),
\label{eq:metadpotential}
\end{equation}

\noindent where $B, \Delta$ and $\tau_{g}$ are constants representing the Gaussian height, width and frequency at which the Gaussians are added, respectively. $\nu(t^{\prime}) = \zeta(\mathbf{r}(t^{\prime}))$ represents the the value of the CV at time $t^{\prime}$, and $t^{\prime}<t$ must be satisfied. If $B$ is large, the free energy surface will be explored at a fast pace, but the reconstructed profile will be affected by large errors. Instead, if $B$ or $\tau_{g}$ are small the reconstruction will be accurate, but it will take a longer time.

The basic assumption of metadynamics is that after a sufficiently long time, $\displaystyle \lim_{t \to \infty} U_{G}(\zeta,t)$ provides an estimate of the free energy landscape, $F(\zeta)$, modulo a constant. However, since the potential of the system changes every time a Gaussian is added, $U_{G}(\zeta,t)$ oscillates around $F(\zeta)$ at long times. Another method that ensures that the FEL converges more smoothly is the Well-Tempered Metadynamics (WTMetaD). In this variant of MetaD, the height of the Gaussians added is no longer a constant but it scales at each timestep, so the bias potential is obtained by replacing $B$ in Eq.~\ref{eq:metadpotential} by:

\begin{equation}
\omega = B\mathrm{exp} \left[ -U_{G}(\zeta(\mathbf{r}(t^{\prime})),t^{\prime})/ \Delta T \right],
\label{eq:WTmetadpotential}
\end{equation}

\noindent where $\Delta T$ has the dimension of temperature. This parameter is related to the bias factor ($\gamma=(T+\Delta T)/T$), which indicates how fast the Gaussian height decreases in a simulation. The higher the value of $\gamma$, the slower the height decrease. Therefore, $\gamma  \to \infty$ represents normal MetaD simulations and $\gamma=1$ represent ordinary molecular dynamics (MD) simulations.

\subsection{The collective variable}
Here we define a collective variable ($d_p$, described in the main text) that can distinguish between planar and non-planar geometries of the DNAns. In practice, $d_p$ is obtained from the direction of each of the dsDNA arms in a nanostar. This is done in the following way: First we compute the position, $\mathbf{r}_{core}$, of the DNAns core. This is done by locating the base-pair (bp) closer to the FJC in each arm (three in total) and calculating the centre of mass (COM) using the position of the six nucleotides previously selected. Then, the unitary vectors ($\uvec{e}_{1},\uvec{e}_{2},\uvec{e}_{3}$) pointing from the core of the molecule to the COM of the bp closer to the sticky end in each arm, define the direction of the three dsDNA arms.

To compute $d_p$ we find two vectors ($\mathbf{A}$ and $\mathbf{B}$) that connect the end of any two arms, for example:

\begin{align}
\mathbf{A} = \uvec{e}_{2} -\uvec{e}_{1}. \nonumber\\
\mathbf{B} = \uvec{e}_{3} -\uvec{e}_{1}.
\end{align}

\noindent These vectors define a plane whose normal can be found as the cross product $\mathbf{n}=\mathbf{A} \times \mathbf{B}$. Then, the distance $d_{p}$ from this plane to the core of the molecule, is given by the projection of any of the three dsDNA arms (for example $\uvec{e}_{2}$) onto the unitary normal vector:

\begin{equation}
d_{p}=\left\vert \uvec{e}_{2} \cdot \uvec{n} \right\vert.
\end{equation}

The range of our CV is from $d_p=0$ (for a completely planar DNAns) to $d_p \lesssim 1$. The larger the value of $d_{p}$ the less planar the molecule. Note that $d_p$ cannot be equal to 1 due to the excluded volume interaction between arms.

\begin{figure*}[htpb]
\centering
\includegraphics[width=1.0\textwidth]{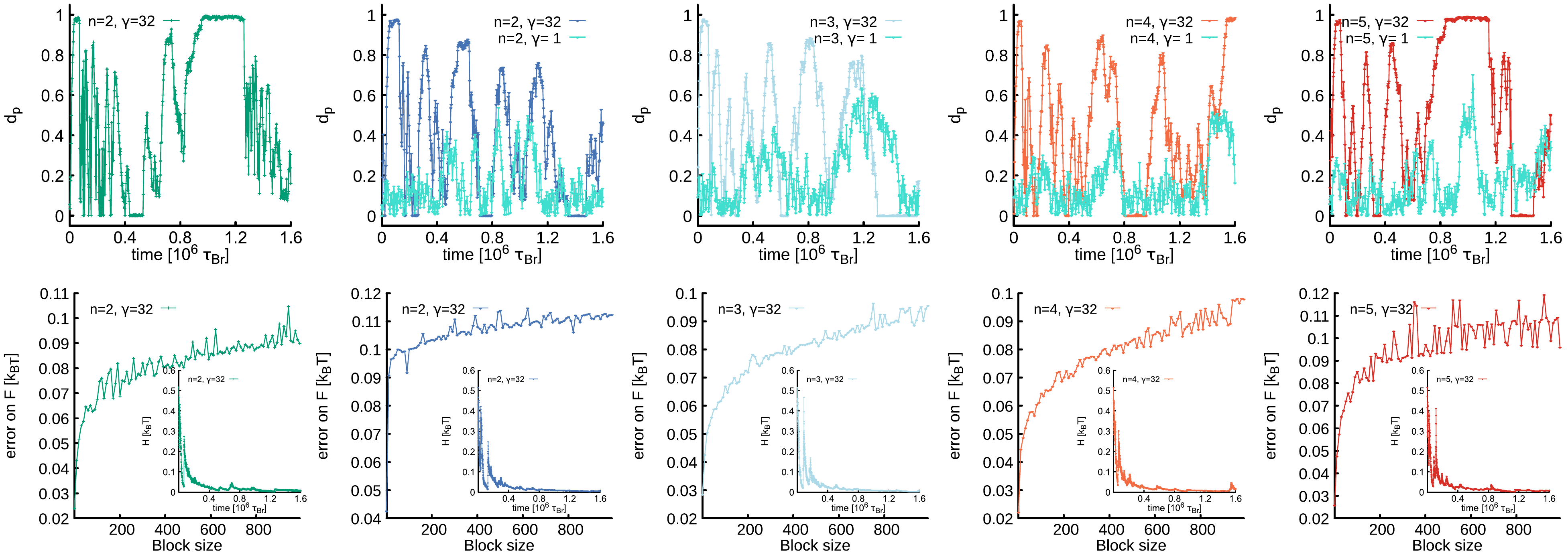}
\caption{Metadynamics simulations. First column shows result at a high salt concentration [NaCl]=1 M for $n=2$ unpaired nucleotides at the FJC. From column two to five we show results at [NaCl]=0.15 M for $n=2,3,4$ and 5, respectively. Top plots show the time evolution of the collective variable ($d_{p}$). In each plot, two curves with different colour are shown. In turquoise we show results for a normal MD simulation without bias ($\gamma=1$) and the other colour represents results from WTMetaD simulations with $\gamma=32$. Bottom depicts the error analysis on $d_{p}$ as function of the block size. Insets show the temporal evolution of the Gaussian Height for the WTMetaD}
\label{fig:metaD}
\end{figure*}
\subsection{Implementation}
Here we use the plumed\cite{plumed} library implemented in LAMMPS~\cite{LAMMPS} (Large Scale Molecular Massively Parallel Simulator) to perform Well Tempered Metadynamics of a three-armed DNA nanostar simulated with the oxDNA model. The sequence of this molecule is given in Table~\ref{M-table:seq} of the main text and it corresponds to the one with $n=2$ unpaired nucleotides at the FJC. We also performed WTMetaD to molecules with $n=3,4$ and 5 (see Fig.~\sref{fig:metaD}). The parameters used in the WTMetaD simulations are: $\gamma=32$, $B=0.5 k_{B}T$, $\Delta=0.05\sigma$ and $\tau_{g}=4\times 10^{3}$ Brownian time-steps. Each simulation was run for a total of $1.6\times 10^{6}$ $\tau_{Br}$.

As described above, by choosing carefully the values of the bias factor ($\gamma$) in the WTMetaD we can compare results from different methods, namely, MD ($\gamma=1$) and WTMetad ($\gamma=32$). Top plots in Fig.~\sref{fig:metaD} show the time evolution of $d_p$. We observe that in simulations without the bias potential (shown in turquoise) $d_{p}$ does not explore the full range of possible values ($d_p$ is defined in the interval [0,1]). The maximum value reached is $\sim 0.65$, it only happens briefly and for $n=5$. Instead, in MetaD the CV is able to explore the whole range of configurations. Bottom plots show the error block analysis of the FEL, which is used to assess the convergence of the WTMetaD simulations. As expected, the error increases with the block size until it reaches a plateau in correspondence of a dimension of the block that exceeds the correlation between data points.


\section{\label{App.detailsCG}Coarse-Grained Molecular Dynamics Simulations}
Molecular dynamics simulations are performed using the model described in the main text. Here we refer to the details on those simulations. The Langevin integration of the system was carried out using LAMMPS~\cite{LAMMPS} (Large Scale Molecular Massively Parallel Simulator) in an NVT ensemble by a standard velocity-Verlet algorithm with integration time-step $\delta t=0.01$. The position of particles in the system obeys the following equation

\begin{equation}
m\frac{d^{2} \mathbf{r}}{dt^{2}}= -\xi \frac{d \mathbf{r}}{dt} - \nabla U + \sqrt{2k_{B}T\xi} \Lambda(t),
\label{eq.langevin}
\end{equation}

\noindent where $m=1$ is the mass of the particle, $\mathbf{r}$ represents its position. $\xi$ is the friction and $\Lambda(t)$ is the white noise term with zero mean which satisfies $\langle \Lambda_{\alpha}(t) \Lambda_{\beta}(s) \rangle = \delta_{\alpha \beta} \delta(s-t)$ along each Cartesian coordinate represented by the Greek letters. The total potential field experience by a particle is $U$ and depends on the interactions set for a particular system.

In our coarse-grained model, each DNA nanostars is represented by a rigid body made up seven beads accounting for its core structure. The excluded volume of beads is modelled via a truncated and shifted Lennard-Jones (LJ) potential:
\begin{equation}
U_{LJ}(r)  =  4\epsilon \left[ 
\left( \frac{\sigma}{r} \right)^{12} 
- \left( \frac{\sigma}{r} \right)^{6} + \frac{1}{4} \right] \, ,
\label{eq:LJ}
\end{equation}
if $r<2^{1/6}\sigma$, and $U_{LJ}(r)=0$ otherwise. Here $\sigma=2.5$ nm represents the diameter of a spherical bead, $\epsilon=1.0$ parametrises the strength of the repulsion and $r$ is the Euclidean distance between the beads.

\begin{figure}[t]
\centering
\includegraphics[width=0.49\textwidth]{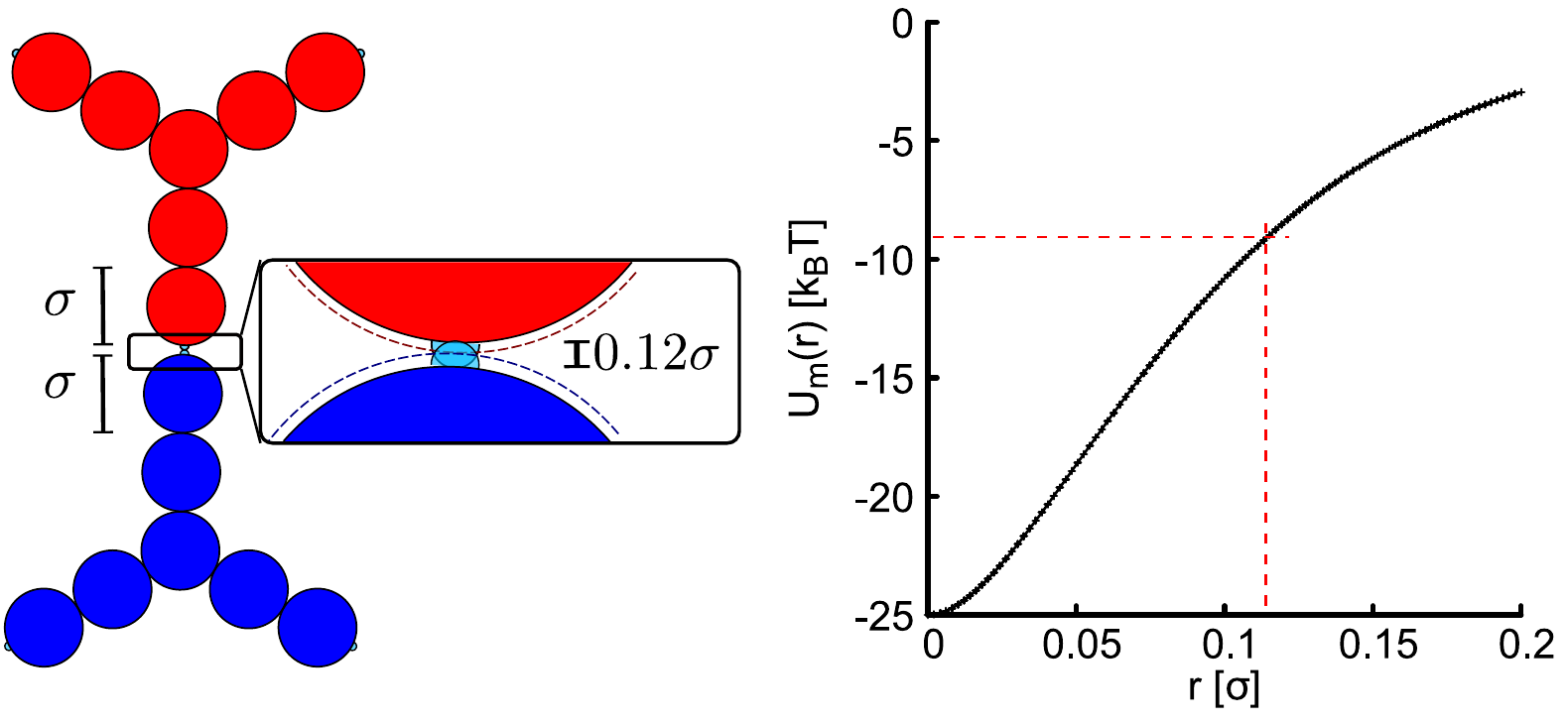}
\caption{Left panel shows schematic representation of two bound DNAns. The hard-sphere repulsion between beads forces patches to remain at a minimum distance of 0.12 $\sigma$. Right panel shows the plot of the morse potential used in simulations to set the attraction between patches (obtained from Eq.~\ref{eq:morse} with $\epsilon_{m}=25.0k_{B}T$, $\alpha_{0}=14 \sigma^{-1}$ and $r_{0}=0$). Although the equilibrium distance between patches is set to zero, the minimum distance allowed by the excluded volume of the beads sets the energy $U_{m}(r=0.12\sigma) \sim 10$ $k_{B}T$ (red dashed lines).}
\label{fig:morse}
\end{figure}

Patches are placed at a distance of $2.5\sigma$ from the core of the molecule and on the surface of the outermost bead along each arm. The sticky ends interaction, responsible for holding two DNAns together, is modelled by a Morse potential:
\begin{equation}
U_{m}(r)  =  \epsilon_{m} \left[ e^{-2\alpha_{0}(r-r_{0})} - 2 e^{-\alpha_{0}(r-r_{0})}\right] \, ,
\label{eq:morse}
\end{equation}
for $r<R_{c}$. Here, $r$ represents the distance between patches of two adjacent nanostars, $r_{0}=0$ is their equilibrium distance and $R_{c}=0.2\sigma$ is the cut-off distance of attraction. The amplitude of the potential is set to $\epsilon_{m}=25.0k_{B}T$ and $\alpha_{0}=14 \sigma^{-1}$ controls the width of the potential. These parameters were chosen to ensure that during the simulation any of the three arms of one ns can hybridize with any (but only one) of the arms of another ns. Importantly, the centre of a patch is placed at the edge of the last bead forming a dsDNA arm, i.e., at a distance of 0.5 $\sigma$ from the centre of the bead. Since the hard-sphere repulsion of beads (set by the LJ potential in Eq.~\ref{eq:LJ} ) covers a radius of $(2^{1/6}\sigma)/2=0.56\sigma$, we expect hybridized patches to be at a minimum distance of $\sim 0.12\sigma$. Therefore, the binding energy felt by hybridized patches is $\sim 10 k_{B}T$ (see Fig.\sref{fig:morse}), in agreement with the activation energy $E_{a}$ mentioned in the main text.

\begin{figure*}[ht]
\centering
\includegraphics[width=0.9\textwidth]{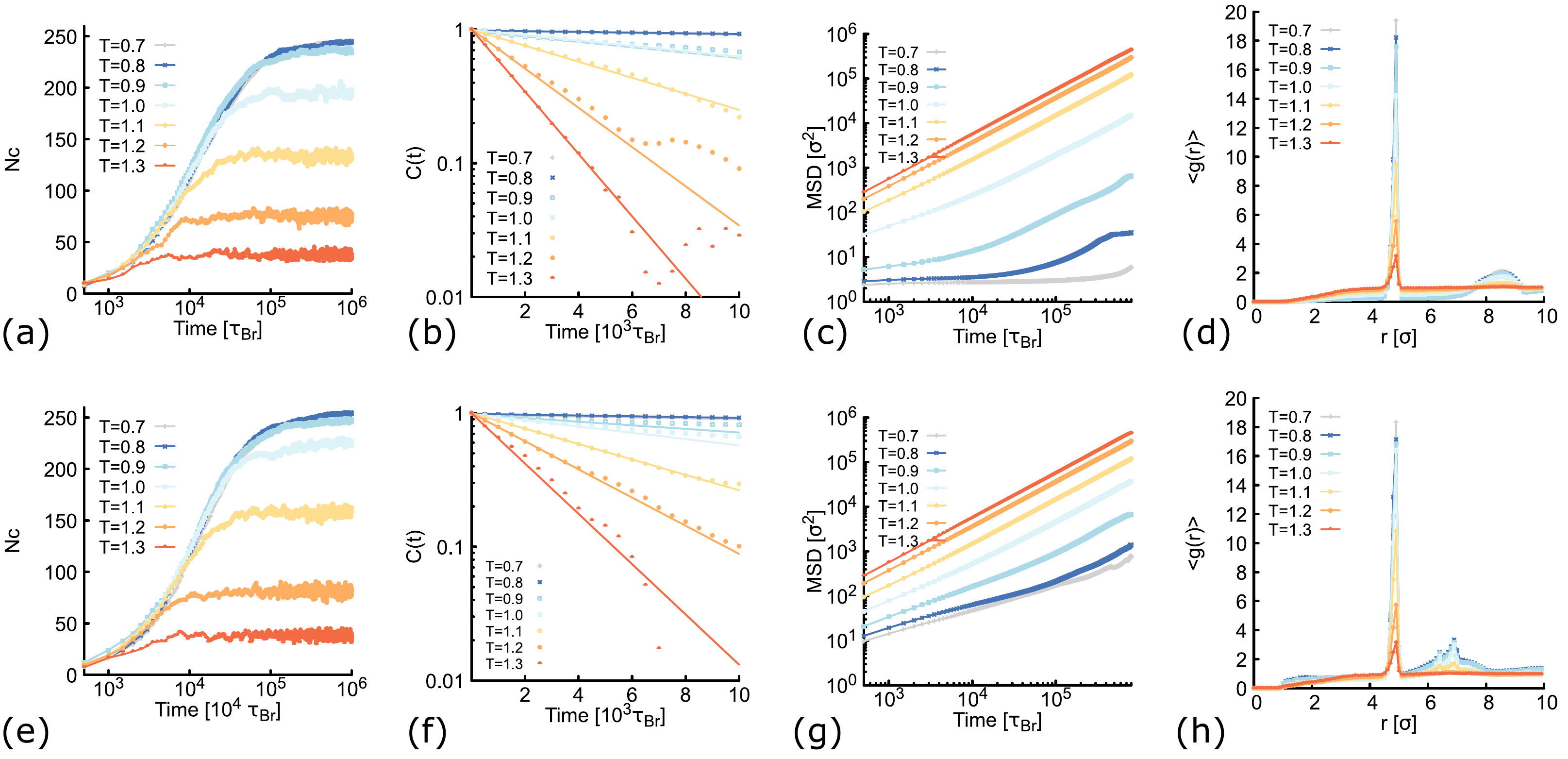}
\caption{Time evolution of different quantities computed from simulations with planar (top panels) and non-planar (bottom panels) molecules. Different colours represent different temperatures (quoted in units of $\epsilon/k_{B}$). (\textbf{a,e}) show examples of trajectories obtained during the network formation. The curves represent the number of contacts $N_{c}$ at a given time t. In order to compute the fraction of connected patches $\vartheta(t)$, we have to multiply $N_{c}$ by the normalization factor $2/Nf$. (\textbf{b,f}) Log-linear plot of the autocorrelation function of $N_{c}(t)$ computed from Eq.\ref{eq:autocorr}. Dots represent results from simulations and lines are a fit to an exponential decay, with decay constant $\tau_{c}$. (\textbf{c,g}) show log-log plots of the MSD averaged over all the molecules in the system. (\textbf{d,h}) Radial distribution function averaged over configurations at long times (when the network is formed).}
\label{fig:changeT}
\end{figure*}

It is worth mentioning here that in the CG model we quote time in units of $\tau_{Br}$, the Brownian time, which is proportional to the time needed for a DNA bead to diffuse its own size. Finally, we note that the total excluded volume ($V_{1}$) of one DNAns is given by the sum of the volume of all its structural beads:
\begin{equation}
V_{1}=7\frac{\pi \sigma^{3}}{6}
\end{equation}

In all our simulations the initial configuration is obtained by placing $N$ nanostars in a cubic simulation box of length $L=40\sigma$ and with periodic boundary conditions. The volume fraction of this system is $\rho=NV_{1}/L^{3}$. We set $\epsilon_{m}=0$ so nanostars cannot hybridized and we equilibrate the system for $5\times10^5$ $\tau_{Br}$. We then turn on the attraction between patches and record the evolution of our system for $10^6$ $\tau_{Br}$. Results for planar and non-planar nanostars at different temperatures are shown in Fig.~\sref{fig:changeT}.

\section{Network formation at fixed concentration}
\subsection*{Melting curve}
In the main text we show the melting profiles $\langle \vartheta\rangle$ as a function of the temperature of the system in Figure~\ref{M-fig:meltandtc}(a). This is obtained by averaging over the last 2 $10^5$ $\tau_{Br}$ of the trajectories obtained for the number of contacts $N_{c}(t)$ (multiplied by the normalization factor $2/Nf$). Examples are shown in Figs.~\sref{fig:changeT}(a,e). It is clear that all the samples start from an equilibrated configuration of unconnected nanostars ($N_{c}(0)=0$); after turning on the attraction between patches the system evolves to a new steady state where a network is formed. The higher the temperature the more unconnected the stars and the smaller the value of $<\vartheta>$.

\subsection*{The autocorrelation function}
From the trajectory of $N_{c}$ at long times, it is also possible to compute the time for network reconfiguration ($\tau_{c}$). This is done by first finding the autocorrelation function of $N_{c}(t)$, which in the discrete case has the form:

\begin{equation}
c(t)= \displaystyle \frac{ \sum_{t=0}^{t_{f}-t'} (N_{c}(t)-\mu) (N_{c}(t+t')-\mu) }{\sum_{t=0}^{t_{f}}(N_{c}(t)-\mu)^{2}},
\label{eq:autocorr}
\end{equation}

\noindent where $t_{f}$ is the total number of time data points, $t'$ is the lag time and $\mu$ is the mean of the data. Results at different temperatures are shown in Figs.~\sref{fig:changeT}(b,f). The autocorrelation function follows an exponential decay $c(t)\propto e^{-t/\tau_{c}}$, from which $\tau_{c}$ is computed. In the main text we show $\tau_{c}$ as function of $T$ in Figure~\ref{M-fig:meltandtc}(b). We also note that, as expected, the data at lower temperatures has larger correlations. This is in agreement with the MSD plots shown in Figs.~\sref{fig:changeT}(c,g), where the system at lower temperatures has a smaller mobility.

\section{Structural analysis}
\subsection*{The radial distribution function}
The structure of our networks is analysed by computing the radial distribution function (RDF) of beads at the core of the DNA nanostars:

\begin{equation}
g(r)=\frac{1}{4\pi \rho_{av}M}\sum_{i=1}^{M}\sum_{j\neq i}^{M} \langle \delta( \vert r_{ij}-r \vert ) \rangle,
\label{eq:rdf}
\end{equation}

\noindent where $\rho_{av}=N/L^{3}$ is the averaged density of core beads in the simulation, $r_{ij}$ represents the distance between the cores of molecules $i$ and $j$. The sum counts the number of DNAns cores that are at a distance $r$.

In each of our simulations there is a total of $N=175$ nanostars (and core beads). We compute $g(r)$ using the wrapped coordinates in the following way: (i) First, we choose the $M$ core beads that are inside an sphere of radius $R_{select}=L/4$ with centre at the origin of the box. (ii) We loop over each one of this $M$ particles, and we count the number of core beads neighbours that are at a distance $r$. This search radius can cover up to a distance $L/4$ and in the count we consider all the core beads in the simulation (not only the $M$ selected in (i)). (iii) We use this information in Eq.~\sref{eq:rdf} to compute $g(r)$. It is worth noting here that we repeat this procedure to compute $g(r)$ in all the configurations in the steady state (at long times and when $\vartheta(t)$ has reached a plateau) and we take the average over these configurations.

Figs.~\sref{fig:changeT}(d,h) show the average RDF obtained at different temperatures for planar and non-planar molecules, respectively. In both cases, $g(r)$ shows a maximum located near to $5\sigma$, corresponding to the average distance between the cores of two bound nanostars. As temperature increases, the hight of this maximum decreases, meaning that there are less contacts between nanostars.

\subsection*{Largest cluster in the network}
\begin{figure}[ht]
\centering
\includegraphics[width=0.5\textwidth]{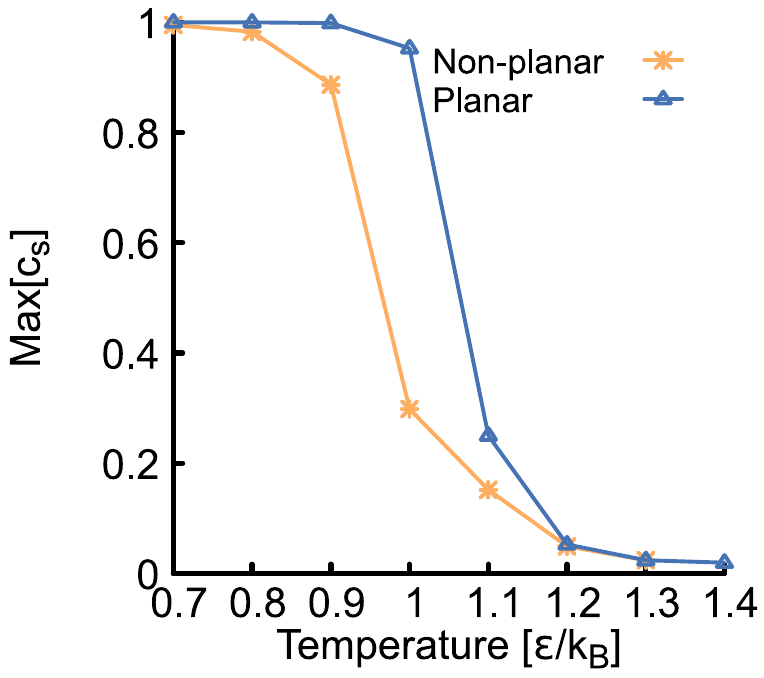}
\caption{Number of connected nanostars in the largest cluster of the network (normalized by the total number of DNAns) as function of the temperature for planar (blue) and non-planar (orange) molecules. All the data shown here correspond to simulations performed at $\rho=0.01$}
\label{fig:cluster}
\end{figure}

Figures~\sref{fig:cluster} shows the size of the largest cluster in the networks made of either, planar or non-planar molecules. Over the whole range of temperatures explored, the size of the largest cluster (Max($C_{s}$)) is larger in the case of networks made out of planar molecules.

\subsection*{Green-Kubo relations}
The autocorrelation function ($G(t)$) is defined as:

\begin{equation}
G(t) = \frac{L^3}{3k_{B}T} \displaystyle \sum_{\alpha \neq \beta} P_{\alpha\beta}(0) P_{\alpha\beta}(t) 
\end{equation}

\noindent where  $P_{\alpha\beta}$ represents the out-off diagonal component ($P_{xy}$, $P_{xz}$ and $P_{yz}$) of the stress tensor. Results from long simulations ($1.6\times 10^{6} \tau_{Br}$) are shown in Fig.~\ref{M-fig:structure}(c) of the main text. The autocorrelation was computed using the multiple-tau correlator method described in reference \cite{taucorr} and implemented in LAMMPS with the \textit{fix ave/correlate/long} command. This computation ensures that the systematic error of the multiple-tau correlator is always below the level of the statistical error of a typical simulation (see LAMMPS documentation).

The viscosity ($\eta$) of the system is then obtained by computing the following integral
\begin{equation}
\eta = \displaystyle \int_{0}^{t \to \infty} G(t) \text{d}t.
\end{equation}

\noindent in practice we compute the numerical discrete integral of the curves in Fig.~\ref{M-fig:structure}(c). We obtained the viscosity in simulation units for planar ($\eta_{p}=573$ $ k_{B}T\tau_{Br}/\sigma^{3}$) and non-planar ($\eta_{np}=29$ $k_{B}T\tau_{Br}/\sigma^{3}$) networks. To convert these results to real units we use $\sigma=2.5$ nm, $k_{B}T=4.11$ $pN\cdot nm$ and the Brownian time:

\begin{equation}
\begin{aligned}
\tau_{Br} &= \sigma^{2}/D \\
          &= \sigma^{2}/\mu_{s}k_{B}T\\
          &= 3\pi\eta_{s} \sigma^{3} / k_{B}T.
\end{aligned}
\end{equation}

\noindent Assuming that the solvent is water (with viscosity $\eta_{s} = 1$ $mPa\cdot s$) we obtain $\tau_{Br} = 36$ ns. Therefore, the viscosities of the two networks map to $\eta_{p} = 5.5 Pa\cdot s$ and $\eta_{np} = 0.28 Pa\cdot s$.

\section{Varying $d_{p}$}
\begin{figure*}[ht]
\centering
\includegraphics[width=1.0\textwidth]{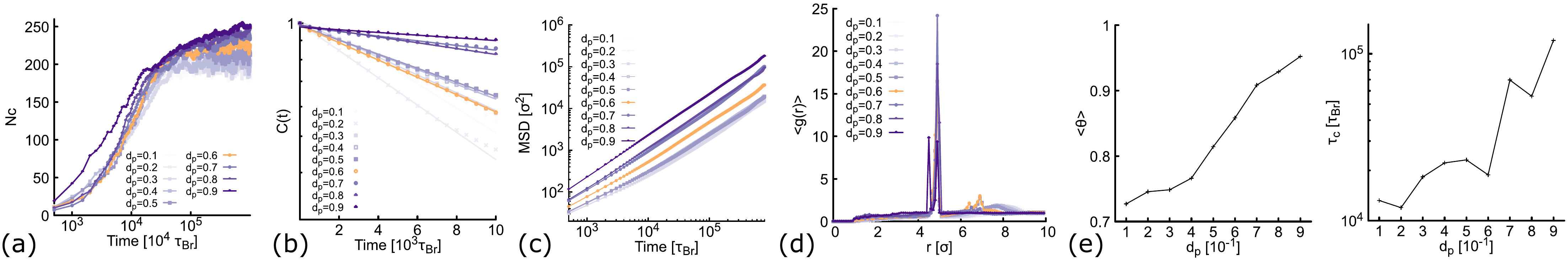}
\caption{Time evolution of different quantities computed from simulations of DNAns with different degree of planarity at $\rho=0.01$ and fixed temperature $T=1.0 \epsilon/k_{B}$. Different colours represent different planarity: from $d_{p}=0.1$ (lightest colour) to $d_{p}=0.9$ (darkest colour). In orange we show results for $d_{p}=0.6$, the one used so far as non-planar in the main text and ESI. (\textbf{a}) shows trajectories obtained during the network formation. The curves represent the number of contacts $N_{c}$ at a given time t. (\textbf{b}) Log-linear plot of the autocorrelation function of $N_{c}(t)$ computed from Eq.\ref{eq:autocorr}. Dots represent results from simulations and lines are a fit to an exponential decay, with decay constant $\tau_{c}$. (\textbf{c}) shows log-log plots of the MSD averaged over all the molecules in the system. (\textbf{d}) Radial distribution function averaged over configurations at long times (when the network is formed). (\textbf{e}) Fraction of contacts ($\vartheta$, left panel) and $\tau_{c}$ (right panel) as function of $d_{p}$.}
\label{fig:changeplanarity}
\end{figure*}

So far we have considered non-planar nanostars with a fixed value of $d_{p}=0.6$. Here, we check that results are consistent for networks formed by DNAns with a different degree of planarity when simulations are performed at a fixed temperature ($T=1.0\epsilon/k_{B}$) and concentration ($\rho=0.01$). Fig.~\sref{fig:changeplanarity}(a-b) show the temporal evolution of $N_{c}(t)$ and $c(t)$. We observe that as $d_{p}$ increases, more contacts between DNAns are formed and the contacts-correlation is larger. The MSD in Fig.~\sref{fig:changeplanarity}(c) shows that as DNAns become less planar, their mobility increases. Therefore, we expect that the larger the value of $d_{p}$ the lower the viscosity of the system. Interestingly, the RDF in Fig.~\sref{fig:changeplanarity}(d) shows that the two local maxima located at $r\in [6,8]$ are developed only for intermediate values of $d_{p}$. Finally, panel (e) shows the change in $\theta$ and $\tau_{c}$ as function of $d_{p}$.

\bibliographystyle{apsrev4-1}
\bibliography{cghydrogel} 

\makeatletter\@input{intermediate1.tex}\makeatother
\makeatletter\@input{intermediate2.tex}\makeatother